\documentstyle[amssymb]{elsart}

\begin{document}

\input epsf

\begin{frontmatter}
\title{Driven Diffusive Systems: \ An Introduction and Recent Developments}
\author{B. Schmittmann and R. K. P. Zia}
\address{Center for Stochastic Processes in Science and Engineering\\
Physics Department\\
Virginia Polytechnic Institute and State University \\
Blacksburg, VA, 24061-0435 USA}
\begin{abstract}
Nonequilibrium steady states in driven diffusive systems exhibit many features which
are surprising or counterintuitive, given our experience with equilibrium systems. 
We introduce the prototype model and review its unusual behavior in different 
temperature regimes, from both a simulational and analytic view point. We then 
present some recent work, focusing on the phase diagrams of driven bi-layer systems
and two-species lattice gases. Several unresolved puzzles are posed.  
\end{abstract} 
\end{frontmatter}

\section{Introduction}

In nature, there are no true equilibrium phenomena, since all of these
require infinite times and infinite thermal reservoirs or perfect
insulations. Nevertheless, for a large class of systems, it is possible to
set up conditions under which predictions from {\em equilibrium} statistical
mechanics provide excellent approximations, as many of the inventions of the
industrial revolution can attest to. By contrast, non-equilibrium phenomena
are not only ubiquitous, but often elude the powers of the Boltzmann-Gibbs
framework. Unfortunately, to date, the theoretical development of {\em %
non-equilibrium} statistical mechanics is at a stage comparable to that of
its equilibrium counterpart in the days before Maxwell and Boltzmann. Using
the intuition developed by studying equilibrium statistical mechanics, we
are often ``surprised'' by the behavior displayed by systems far from
equilibrium, even if they appear to be in time-independent states. For
example, the well honed arguments, based on the competition between energy
and entropy, frequently fail dramatically. In these lectures, we will
present some explorations into the intriguing realm of {\em non-equilibrium
steady states}, focusing only on a small class, namely, driven diffusive
systems.

Motivated by both the theoretical interest in non-equilibrium steady states
and the physics of fast ionic conductors \cite{FIC}, Katz, Lebowitz and
Spohn \cite{KLS} introduced a deceptively minor modification to a well-known
system in equilibrium statistical mechanics: the Ising \cite{Ising} model
with nearest-neighbor interactions. In lattice gas language \cite
{YangLee}, the time evolution of this model can be specified by particles
hopping to nearest vacant sites, with rates which simulate coupling to a
thermal bath as well as an external field, such as gravity. If ``brick
wall'' boundary conditions are imposed in the direction associated with
gravity (particles reflected at the boundary, comparable to a floor or a
ceiling), and if the rates obey detailed balance, then this system will
eventually settle into an {\em equilibrium} state, similar to that of gas
molecules in a typical room on earth. Although there is a local bias in the
hopping rates (due to gravity), thermal equilibrium is established by an
inhomogeneous particle density, at all $T<\infty $. However, if {\em periodic%
} boundary conditions are imposed, then translational invariance is
completely restored so that, in the final steady state, the particle density
is {\em homogeneous}, for all $T$ above some finite critical $T_{c}$, while
a current will be present. Clearly, for gravity, such boundary conditions
can be imposed only in art, as by M.C. Escher \cite{MCE}. In physics, it is
possible to establish such a situation with an {\em electric} field, e.g.,
by placing the $d=2$ lattice on the surface of a cylinder and applying a
linearly increasing magnetic field down the cylinder xis. If the particles
are charged, they will experience a local bias everywhere on the cylinder
and a current will persist in the steady state. Echoing the physics of fast
ionic conductors, we will therefore use the term ``electric'' field to
describe the external drive and imagine our particles to be ``charged'', in
their response to this drive. This is the prototype of a ``driven diffusive
system.'' Now, such a system constantly gains (loses) energy from (to) the
external field (thermal bath), so that the time independent state is by no
means an equilibrium state, \`{a} la Boltzmann-Gibbs. Instead, it is a {\em %
non-equilibrium steady state}, with an unknown distribution in general. As
discovered in the last decade, modifying the Ising model to include a simple
local bias leads to a large variety of far-from-simple behavior. The scope
of these lectures necessarily limits us to only a bird's eye view. Since a
more extensive review has been published recently \cite{DL17}, we will
restrict these notes to a brief ``introduction'' to this subject. Instead,
we choose to include some developments since that review.

For completeness, we summarize, in Section II, the lattice model introduced
by KLS and the Langevin equation believed to capture its essence in the
long-time large-distance regime. The next section (III) is devoted to some
of the surprising behavior displayed by this system, at temperatures far
above, near, and well below $T_{c}$. A collection of recent developments is
presented in Section IV. In the final section (V), we conclude with a brief 
summary.

\section{The Microscopic Model and a Continuum Field Approach}

On a square lattice, with $L_{x}\times L_{y}$ sites and toroidal boundary
conditions, a particle or a hole may occupy each site. A configuration is
specified by the occupation numbers $\{n_{i}\}$, where $i$ is a site label
and $n$ is either 1 or 0. Occasionally, we also use spin language, defining $%
s\equiv 2n-1=\pm 1$. To access the critical point, half-filled lattices are
generally used in Monte Carlo simulations: $\sum_{i}n_{i}=$ $L_{x}L_{y}/2$.
The particles are endowed with nearest-neighbor attraction (ferromagnetic,
in spin language), modeled in the usual way through the Hamiltonian: $%
H=-4J\sum_{<i,j>}n_{i}n_{j}$, with $J>0$. The external drive, with strength $%
E$ and pointing in the $-y$ direction, will bias in favor of particles
hopping ``downwards''. To simulate coupling to a thermal bath at temperature 
$T$, the Metropolis algorithm is typically used, i.e., the contents of a
randomly chosen, nearest-neighbor, particle-hole pair are exchanged, with a
probability $\min [1,e^{-(\Delta H-\epsilon E)/k_{B}T}]$. Here, $\Delta H$
is the change in $H$ after the exchange and $\epsilon =(-1,0,1)$, for a
particle attempting to hop (against, orthogonal to, along) the drive. Note
that these dynamic rules conserve particle number. For $E=0$, this system
will eventually settle into an equilibrium state which is precisely the
static Ising model \cite{Ising,YangLee}. In the thermodynamic limit, it is
known to undergo a second order phase transition at the Onsager critical
temperature $T_{c}(0)=(2.2692..)J/k_{B}$. When driven ($E\neq 0$), this
system displays the same qualitative properties, i.e., there is a disordered
phase for large $T$, followed by a second order transition into a
phase-segregated state for low $T$. However, with more scrutiny, this
superficial similarity gives way to puzzling surprises. In particular, $%
T_{c}(E)$ appears to be monotonically {\em increasing} with $E$, saturating
at about $1.41T_{c}(0)$ \cite{KTLJSW} for $E\gg J$. Why should $T_{c}(\infty
)>T_{c}(0)$ be surprising? Consider the following ``argument''. For very
large $E$, hopping along $y$ becomes like a random walk, in that $\Delta H$
is irrelevant for the rates. Therefore, hops along $y$ might as well be
coupled to a thermal bath at {\em infinite} $T$ (apart from the bias).
Indeed, recall that our system gains energy from $E$ and loses it to the bath,
so that any drive may be thought of as a coupling to a second reservoir with
higher temperature. Then, it seems reasonable that $T$ must be {\em lowered}
to achieve ordering, since this extra reservoir pumps in a higher level of
noise, helping to {\em disorder} the system! To date, there is neither a
convincing argument nor a computation which predicts the correct {\em sign}
of $T_{c}(E)-T_{c}(0)$, let alone the magnitude. In the next Section, we
will briefly review other puzzling discoveries, only a few of which are
understood.

To understand collective behavior in the long-time and large-scale limit, we
often rely on continuum descriptions, which are hopefully universal to some
extent. Successful examples include hydrodynamics and Landau-Ginzburg
theories. Certainly, the $\varphi ^{4}$ theory, enhanced by renormalization
group techniques, offers excellent predictions for both the statics and the
dynamics near equilibrium of the Ising model. Following these lines, we
formulate a continuum theory for the KLS model, in arbitrary dimension $d$.
In principle, such a description can be obtained by coarse-graining the
microscopic dynamics \cite{Eyink} but we will pursue a more phenomenological
approach here. Seeking a theory in the long-time, large wavelength limit, we
first identify the slow variables of the theory. These are typically

\begin{itemize}
\item  ordering fields which experience critical slowing down near $T_{c}$, and

\item  any conserved densities.
\end{itemize}

\noindent The KLS model is particularly simple since it involves a single
ordering field, namely the local ``magnetization'' or excess particle
density, $\varphi ({\bf x},t)$, which is also the only conserved quantity.
Here, ${\bf x}$ stands for $(x_{1},x_{2},...,x_{d-1},x_{d}=y)$. The last
entry denotes the one-dimensional ``parallel'' subspace selected by $E$.
Thus, we begin with a continuity equation, $\partial _{t}\varphi +\nabla 
{\bf j}=0$, and postulate an appropriate form for the current ${\bf j}$. In
the absence of $E$, it simply takes its Model B \cite{hh} form, ${\bf j}(%
{\bf x},t)=-\lambda \nabla \frac{\delta {\cal H}}{\delta \varphi }+{\bf \eta
(x},t)$, with the Landau-Ginzburg Hamiltonian ${\cal H=}\int \{\frac{1}{2}%
(\nabla \varphi )^{2}+\frac{\tau }{2}\varphi ^{2}+\frac{u}{4!}\varphi ^{4}\}$%
. As usual, $\tau \propto T-T_{c}$ and $u>0$. While the first contribution
to ${\bf j}$ is a deterministic term, reflecting local chemical potential
gradients, the second term, ${\bf \eta }$, models the thermal noise. The
noise is Gaussian distributed, with zero mean and positive second moment
proportional to the unit matrix, i.e., $<\eta _{i}\eta _{j}>=2\sigma \delta
_{ij}\delta ({\bf x}-{\bf x}^{\prime })\delta (t-t^{\prime })$, $i,j=1,...,d$%
. In the presence of $E$, we should expect at least two modifications to $%
{\bf j}$, namely first, an additional contribution ${\bf j}_{E}$ modeling
the nonvanishing mass transport through the system and second, the
generation of anisotropies, since $E$ singles out a specific direction. By
virtue of the excluded volume constraint, the ``Ohmic'' current ${\bf j}_{E}$
vanishes at densities $1$ and $0$, corresponding to $\varphi =\pm 1$.
Writing ${\cal E}$ for the coarse-grained counterpart of $E$, pointing along
unit vector ${\bf \hat{y}}$, the simplest form is therefore ${\bf j}_{E}=%
{\cal E}(1-\varphi ^{2})[1+O(\varphi )]{\bf \hat{y}}$, where the $O(\varphi )
$ corrections will turn out to be irrelevant for universal properties. Next,
we consider possible anisotropies. Clearly, all $\nabla ^{2}$ operators
should be split into components parallel ($\partial ^{2}$) and transverse ($%
\nabla _{\bot }^{2}$) to $E$, accompanied by different coefficients. For
example, the anisotropic version of the Model B term $\tau \nabla
^{2}\varphi $ will read $(\tau _{\Vert }\partial ^{2}+$ $\tau _{\bot }\nabla
_{\bot }^{2})\varphi $, with two different ``diffusion'' coefficients for
the parallel and transverse subspaces. Should we expect both of these to
vanish as $T$ approaches $T_{c}(E)$? Or just one of them - but which one?
Recalling that the lowering of $\tau $, in the equilibrium system, is a
consequence of the presence of interparticle interactions, we argue that $E$%
, stirring parallel jumps much more effectively than transverse ones, should 
{\em counteract} this effect in the parallel direction, having less of an
impact in the transverse subspace. Thus, we anticipate that generically $%
\tau _{\Vert }>\tau _{\bot }$, so that criticality, in particular, is marked
by $\tau _{\bot }$ vanishing at positive $\tau _{\Vert }$. This is borne out
by the structure of typical ordered configurations, namely, single strips
aligned with $E$, indicating that ``antidiffusion'', i.e., $\tau _{\bot }<0$%
, dominates in the transverse directions below $T_{c}$. Finally, the drive
also induces anisotropies in the noise terms so that the second moment of
their distribution should be taken as $<\eta _{i}\eta _{j}>=2\sigma
_{i}\delta _{ij}\delta (x-x^{\prime })\delta (t-t^{\prime })$. Since the
transverse subspace is still fully isotropic, we define $\sigma
_{1}=...=\sigma _{d-1}\equiv \sigma _{\Vert }$. Generically, however, we
should expect $\sigma _{d}\equiv \sigma _{\bot }\neq $ $\sigma _{\Vert }$.
Summarizing, we write down the full Langevin equation: 
\begin{eqnarray}
\partial _{t}\varphi ({\bf x},t) &=&\lambda \{(\tau _{\bot }-\nabla _{\bot
}^{2})\nabla _{\bot }^{2}\varphi +(\tau _{\Vert }-\partial ^{2})\partial
^{2}\varphi -2\alpha _{\times }\partial ^{2}\nabla _{\bot }^{2}\varphi + 
\nonumber \\
&&+\frac{u}{3!}(\nabla _{\bot }^{2}+\kappa \partial ^{2})\varphi ^{3}+{\cal E%
}\partial \varphi ^{2}\}-\nabla {\bf \eta (x},t)\qquad ,  \label{lang}
\end{eqnarray}
with noise correlations
\[
<\eta _{i}({\bf x},t)\eta _{j}({\bf x}^{\prime },t^{\prime })>=2\sigma
_{i}\delta _{ij}\delta (x-x^{\prime })\delta (t-t^{\prime })\qquad .
\]

This equation forms the basis for the analytic study of the KLS model.
Fortunately, we need not be concerned about the detailed dependence of its
coefficients on the microscopic $T$, $J$, and $E$. While they are in
principle calculable within an explicit coarse-graining scheme, the
properties that we already outlined above suffice for our purposes.
 Moreover, it turns out that we can simplify the most general form, (\ref
{lang}), depending on the temperature regime considered. We will return to
this discussion in Sections 3.1 and 3.3 below.

To conclude this Section, let us take a glance at the structure of (\ref
{lang}). Its basic form is $\partial _{t}\varphi ={\Bbb F(\varphi },\nabla
\varphi ,...)+\zeta $, with noise correlations $<\zeta ({\bf x},t)\zeta (%
{\bf x}^{\prime },t^{\prime })>=2\sigma {\Bbb N}\delta ({\bf x}-{\bf x}%
^{\prime })\delta (t-t^{\prime })$ and $\sigma >0$ just a positive constant.
For simplicity, we restrict ourselves to a scalar order parameter $\varphi $
and noise ``matrices'' ${\Bbb N}$ which do not depend on $\varphi $ (more
general cases are discussed in, e.g., \cite{risken} and \cite{hkj}). ${\Bbb F%
}$ is a functional of $\varphi $ and its derivatives. Given this basic form,
a steady-state solution $P^{*}[\varphi ]$ for the configurational
probability is easily found \cite{FP} {\em provided} ${\Bbb F}$ is
``Hamiltonian'', i.e., if it can be written as ${\Bbb N}$ acting on a total
functional derivative: ${\Bbb F=-N}\frac{\delta {\cal H}}{\delta \varphi }$.
In this case, $P^{*}$ is simply proportional to $\exp (-{\cal H}/\sigma 
{\cal )}$, irrespective of the choice of ${\Bbb N}$. We will refer to such a
dynamics, in an operational sense \cite{DL17}, as ``satisfying the
fluctuation-dissipation theorem'' (FDT) \cite{FDT}. Model B falls into this
category, with ${\Bbb N=-\nabla }^{2}$. In contrast, for our driven system, $%
{\Bbb F}$ is just the expression in the $\{...\}$ brackets of (\ref{lang})
and $\varsigma \equiv -\nabla {\bf \eta }$, so that ${\Bbb N=-(\sigma }%
_{\Vert }\partial ^{2}+\sigma _{\bot }\nabla _{\bot }^{2})$. Thus, this $%
{\Bbb F}$ is clearly {\em not} Hamiltonian! In this fashion, our continuum
theory reflects the fact that we are dealing with a generic non-equilibrium
steady state. For completeness, we should point out that certain
microscopically non-Hamiltonian dynamics may {\em become} Hamiltonian if
viewed on sufficiently large length scales. The discussion of these
subtleties \cite{GYH,FDT subtle,BS}, while intriguing, is beyond the scope
of these lectures.

\section{Surprising Singular Behavior, near and far from Criticality}

For the $d=2$ Ising model in equilibrium, thermodynamic quantities are
typically analytic, except at a single point, i.e., $T_{c}$. In particular,
being a model with only short-range interactions, correlation functions are
short-ranged far above $T_{c}$, decaying exponentially with distance and
controlled by a finite correlation length. Far below criticality, a half
filled system in a square geometry will display two strips of equal width,
as a result of the co-existence of a particle rich (dense) phase with
hole-rich one. Correlations of the fluctuations within each phase are also
short-ranged. More interestingly, the interfaces between the phases do
represent soft degrees of freedom, being the Goldstone mode of a broken
translational invariance \cite{Goldstone}. One consequence is a divergent
structure factor (Fourier transform of the two-point correlation, of
deviations from being straight). Although this behavior is singular, it is
well understood and can be related to that of a simple random walk \cite{RW}%
. Only at criticality does the system possess non-trivial singular
properties, the {\em nature }of which became transparent only in the 
70's\cite{RG} despite Onsager's tour-de-force in 1944\cite{Onsager}.

When driven into {\em non-equilibrium steady states}, this picture changes
dramatically. In particular, non-trivial singular behavior appears at all $T$%
. While the transition itself remains second order, its properties fall into
a non-Ising class. Here, we will present some of these surprising features
briefly, referring the reader to \cite{DL17} for more details.

\subsection{$T$ far above $T_c$}

Focusing on the behavior far above criticality, let us study the two-point
correlation $G({\bf r})$ (${\bf r=}(x,y)$) and its transform, the structure
factor, $S({\bf k})$. We first point out that the drive clearly induces an
asymmetry between $x$ and $y$, i.e., anisotropy beyond that due to the
lattice. Thus, we should not be surprised if the familiar Ornstein Zernike
form of $S$ ($\propto (1+\xi ^{2}{\bf k}^{2})$) were to become elliptical
rather than circular, e.g., similar to Fig.1a. However, simulations \cite{KLS}
showed that there is a {\em discontinuity} singularity in $S$ at the origin,
as in Fig.1b. To be more precise, we may define 
\begin{equation}
R\equiv \frac{\lim_{k_{y}\rightarrow 0}S(0,k_{y})}{\lim_{k_{x}\rightarrow
0}S(k_{x},0)}
\end{equation}
and measure the discontinuity by $R-1$. Though $R$ does approach unity for $%
T\rightarrow \infty $, it is about 4, for large $E$, even at twice the
critical temperature. Further, it diverges as $T\rightarrow T_{c}$! We
should remind the reader that, for the equilibrium case, $S$ may diverge at
criticality, but $R$ is unity always.

\begin{figure}[tbph]
\hspace*{1.5cm} \epsfxsize=4in \epsfbox{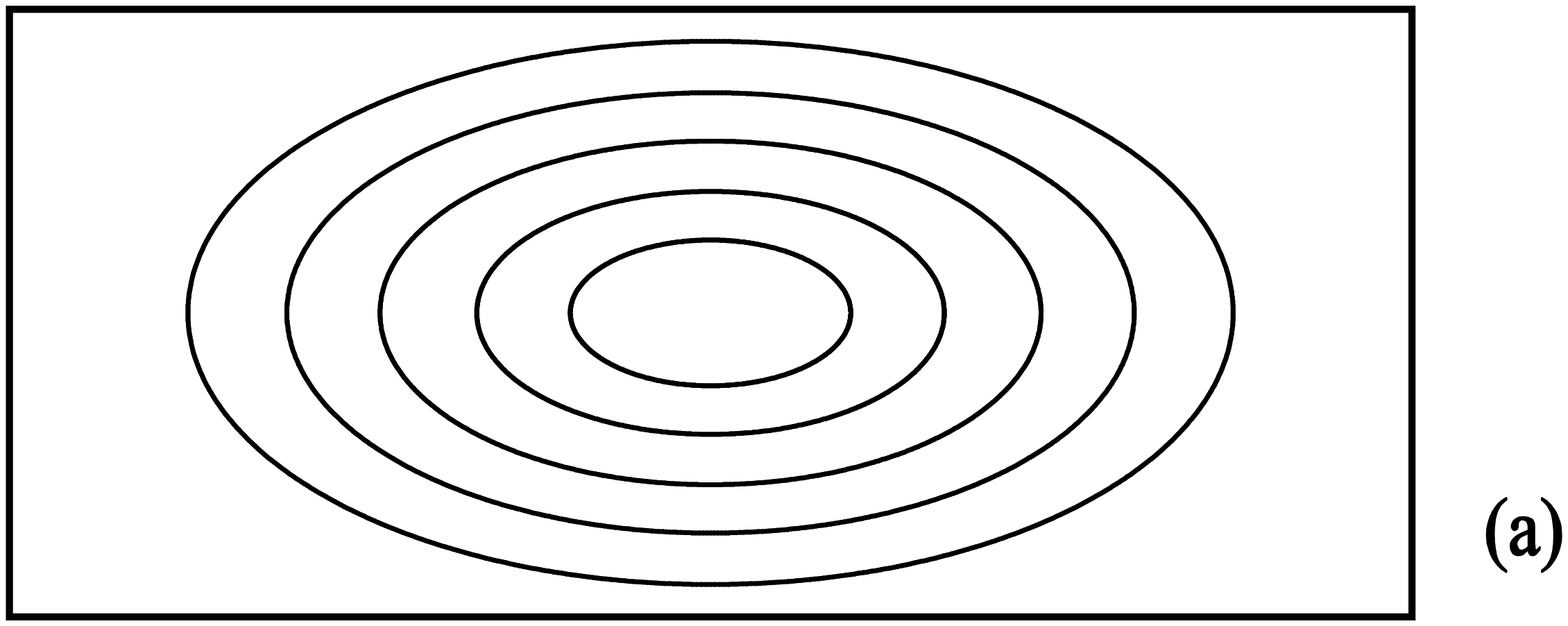}
\par
\vspace*{-3cm} \hspace*{1.5cm} \epsfxsize=4in \epsfbox{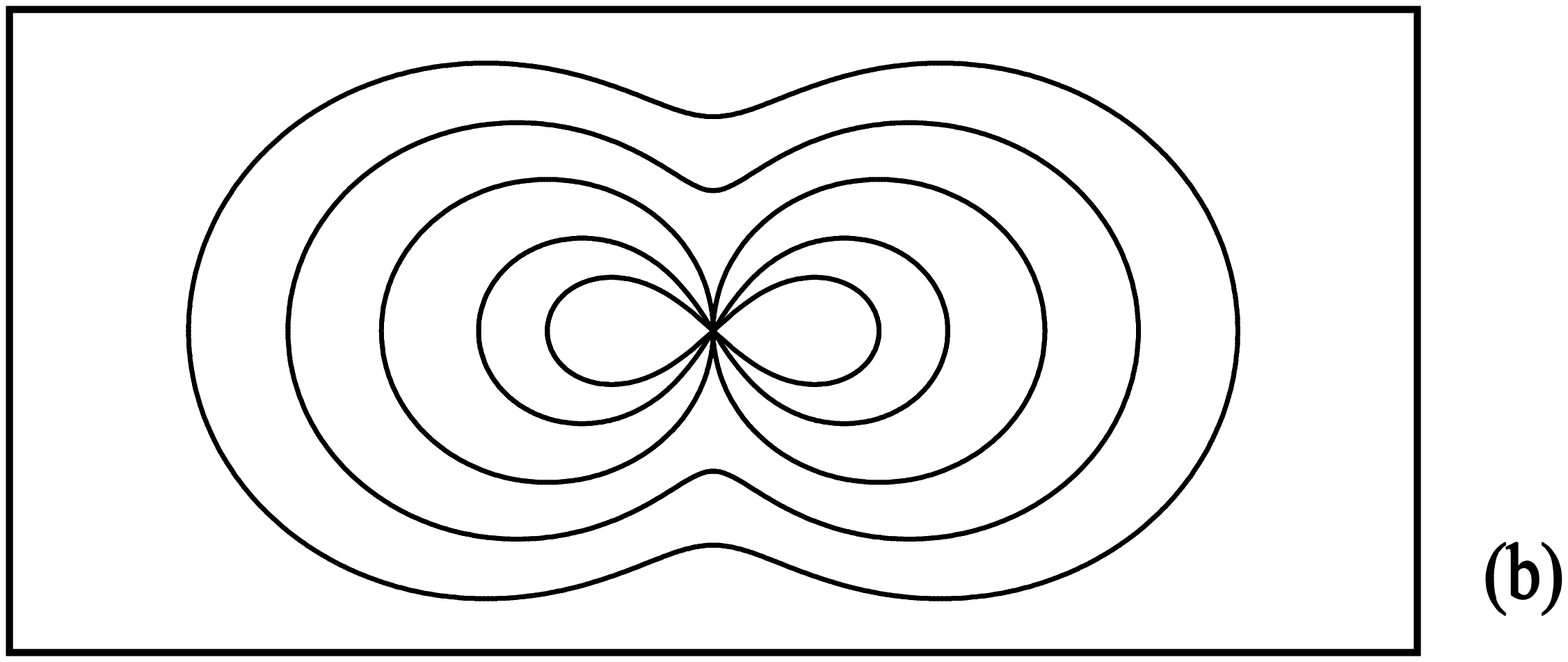}
\vspace*{-2.5cm}
\caption{Schematic plot of structure factors. (a) Ellipses for a typical
anisotropic Ising model in equilibrium. (b) ``Butterfly'' or ``owl'' pattern
for a driven system.}
\end{figure}

Strange as it may seem, this behavior can be understood within the context
of the continuum approach, Eqn. (\ref{lang}). For $T\gg T_{c}$, {\em both} $%
\tau _{\Vert }$ {\em and} $\tau _{\bot }$ are positive. Thus, the local
magnetization $\varphi ({\bf x},t)$ fluctuates around a stable minimum at
zero, and neither the fourth-order derivatives nor the nonlinearities are
necessary for stability. As a consequence, the behavior of the system in the
disordered phase can be described by a much simpler, {\em linear} Langevin
equation, namely, 
\begin{equation}
\partial _{t}\varphi ({\bf x},t)=\lambda ( \tau _{\bot }\nabla _{\bot
}^{2}+\tau _{\Vert }\partial ^{2} ) \varphi - \nabla {\bf \eta }.
\end{equation}
With the help of a Fourier transform, we can easily find $\varphi ({\bf k},t)
$ for any realization of ${\bf \eta }$, and then compute averages over the
Gaussian distributed ${\bf \eta }$, with its second moment given by (\ref
{lang}). This forms the starting point for the discussion of the disordered
phase, quantified by, e.g., equal-time structure factors or three-point
functions \cite{3pf}.

Here, we focus on the former, $S({\bf k})\equiv <\varphi ({\bf k},t)\varphi (-%
{\bf k},t)>$, which can be easily computed: 
\begin{equation}
S({\bf k})=\frac{\sigma _{\bot }k_{x}^{2}+\sigma _{\Vert }k_{y}^{2}}{\tau
_{\bot }k_{x}^{2}+\tau _{\Vert }k_{y}^{2}+O(k_{x}^{4},k_{x}^{2}k_{y}^{2}{\bf %
,}k_{y}^{4}{\bf )}},
\end{equation}
so that $R=\left( \sigma _{\Vert }\tau _{\bot }\right) /\left( \sigma _{\bot
}\tau _{\Vert }\right) $. In order for this $S$ to reduce to the equilibrium
Ornstein-Zernike form, the FDT has to be invoked, which constrains $R$ to
unity. For the driven case, $R$ is no longer constrained, so that a
discontinuity develops. One consequence of such a singularity in $S$ is
that $G$ becomes {\em long-ranged}, decaying as ${\bf r}^{-2}$ at large $|%
{\bf r|}$. The amplitude, in addition to being $\propto (R-1)$, has a
dipolar angular dependence, so that an appropriate angular average of $G$ is
again short-ranged \cite{DL17}.

To end this discussion, let us follow G. Grinstein's argument \cite{GG} that
such long-ranged power law decays {\em should be expected}. Starting with
the dynamic two-point correlation $G({\bf r},t)$, we know that the
conservation law leads to the autocorrelation: $G({\bf 0},t)\rightarrow
t^{-d/2}$. In addition, being a diffusive system, we should expect ${\bf r}%
\sim \sqrt{t}$, at least far from criticality. Using {\em naive }scaling, we
would write $G({\bf r},0)\rightarrow |{\bf r}|^{-d}$! In other words, the
generic behavior of $G$ in a diffusive system is $|{\bf r}|^{-d}$ decay, 
{\em not exponential}. In this case, the equilibrium system is the
non-generic one, in which the amplitude of this generic term is forced to
vanish, by FDT. Our familiarity with equilibrium systems is so strong that,
on first sight, power law decays far above $T_{c}$ appear quite surprising.
Finally, note that a scaling argument of this kind cannot produce the
angular dependence, a crucial feature of the generic singularities of driven
diffusive systems.

\subsection{$T$ far below $T_c$}

Next, we turn to systems far below $T_{c}$. Due to the conservation law,
they typically display phase co-existence, so that the dominant fluctuations
and slow modes are those associated with the interface between the phases.
Again, for the $d=2$ Ising model in equilibrium, there is a wealth of
information on the properties of the interface \cite{eqInterface}. In
particular, being a {\em one}-dimensional object, the interface exhibits
behavior identical to a simple random walk \cite{RW} at the large scales,
such as widths diverging as $\sqrt{L}$. In our case, if we study an
interface aligned along the $y$ axis, then, to a good approximation, we may
specify the configuration by its position along $x$ by the ``height''
function $h(y)$. Known as capillary waves \cite{cap-wav}, these fluctuations
have a venerable history. Since the interface is a manifestation of broken
translational invariance, $h(y)$ are the soft Goldstone modes \cite
{Goldstone}, so that the associated structure factor $S_{h}(q)\equiv
\left\langle h(q)h(-q)\right\rangle $ diverges as $q^{-2}$. Of course, this
property is just the Fourier version of the $\sqrt{L}$ divergence: $%
\int_{1/L}q^{-2}dq=O(L)$. Interfaces with divergent widths are also known as
``rough''; only for $d>2$ may interfaces in crystalline Ising-like models
display transitions from rough to smooth phases.

When {\em driven}, however, the interface width appears to approach a {\em %
finite width} at large distances. In particular, for $E=2J$, using lattices
up to $L_{y}=60$ and plotting the widths vs. $L^{p}$ with various values of $%
p$, we found that the curves did not straighten out with $p$ as low as 0.05 
\cite{LMVZ}. As a phenomenon, roughness suppression is not novel, gravity
being the most common example. However, the drive here is {\em parallel} to
the interface, reminiscent of wind driving across a water surface, which
has a {\em destabilizing} effect by contrast. Subsequently, a more detailed
simulation study \cite{LZ93} of the interface with $L_{y}\leq 600$ showed
that $S(q)\sim q^{-2}$ for large $q$, crossing over to $q^{-0.67}$ for small 
$q$ (Fig. 2). Since $\int_{1/L}q^{-0.67}dq=O(1)$ , the small $q$ behavior is
consistent with $p=0$. Though $0.67$ appears temptingly close to a simple
rational: $2/3$, there is no viable theory so far (despite two valiant
attempts \cite{L88Y93}). Finally, let us note that the crossover from rough
to smooth occurs at the scale of $q\sim E$. Although no precise Monte Carlo
analysis of this crossover has been performed, it is consistent with $E$
having the units of 1/length. In this respect, such a length is comparable
to the capillary length which controls the crossover in the gravitationally
stabilized interface. Of course, in that case, the small $q$ behavior is
simply $q^{0}$!

\begin{figure}[htbp]
\hspace*{1.5cm} 
\epsfxsize=4in \epsfbox{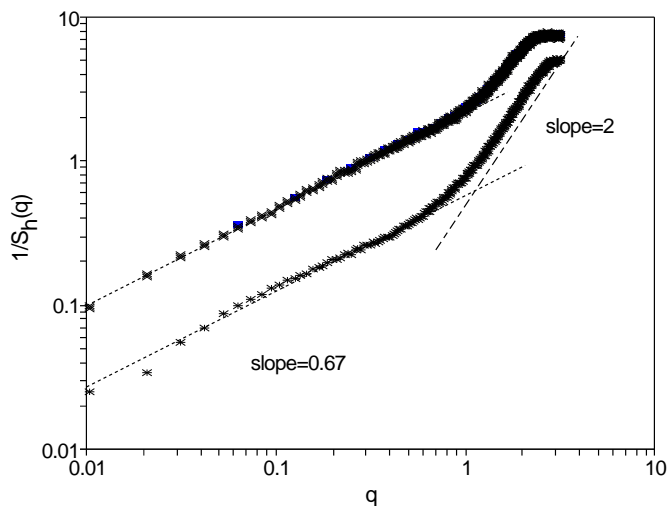} \vspace*{-1cm}
\caption{Log-log plot of interface structure factor vs. wavevector.}
\end{figure}

\subsection{Critical Properties}

Finally, we turn to the critical region, described by $\tau _{\bot }\gtrsim 0
$ in (\ref{lang}). In contrast to the situation for $T\gg T_{c}$, we now
need fourth-order terms in $\nabla _{\bot }$ to stabilize the system against
large-wavelength fluctuations. Similarly, the nonlinear terms are required
to ensure a stable ordered phase below $T_{c}$. However, we still have $\tau
_{\Vert }>0$, so that fourth-order {\em parallel} gradients
may safely be neglected. Thus, near criticality, the leading terms on the
right hand side of (\ref{lang}) are $(\nabla _{\bot }^{2})^{2}\varphi $ and $%
\partial ^{2}\varphi $, implying that parallel and transverse wave vectors, $%
{\bf k}_{\bot }$ and ${\bf k}_{\Vert }$, scale naively as $|{\bf k}_{\Vert
}|\sim |{\bf k}_{\bot }|^{2}$, i.e., parallel gradients are less relevant
than transverse ones. More systematically, a naive dimensional analysis \cite
{ft} reveals that the nonlinearity associated with ${\cal E}$ is the most
relevant one, having an upper critical dimension $d_{c}=5$, distinct from
the usual value of $4$ for the Ising model. Dropping irrelevant terms, (\ref
{lang}) simplifies to 
\begin{equation}
\partial _{t}\varphi ({\bf x},t)=\lambda \{(\tau _{\bot }-\nabla _{\bot
}^{2})\nabla _{\bot }^{2}\varphi +\partial ^{2}\varphi +\frac{u}{3!}\nabla
_{\bot }^{2}\varphi ^{3}+{\cal E}\partial \varphi ^{2}\}-\nabla _{\bot }{\bf %
\eta }_{\bot } ( {\bf x},t)  \label{laec}
\end{equation}
with noise term ($i,j=1,...,d-1$)
\[
<\eta _{\bot i}({\bf x},t)\eta _{\bot j}({\bf x}^{\prime },t^{\prime
})>=2\sigma _{\bot }\delta _{ij}\delta (x-x^{\prime })\delta (t-t^{\prime
}) .
\]
Here, we have rescaled $\tau _{\Vert }$ to $1$ and have kept the (naively
irrelevant) nonlinearity associated with $u$, to ensure a stable theory
below $T_{c}$. This is the starting point for the analysis of universal
critical behavior.

In this regime, large fluctuations on all length scales dominate the
behavior of the system, so that renormalization group techniques are
indispensable. To summarize very briefly, the Langevin equation (\ref{laec})
is recast as a dynamic functional \cite{dyn fun}, followed by a renormalized
perturbation expansion in $\epsilon \equiv d-d_{c}$ \cite{ft}. The quartic
coupling $u$ must be treated as a dangerously irrelevant operator.
Gratifyingly, the series for the critical exponents can be summed, so that
we obtain {\em quantitatively }reliable values even in two dimensions. The
details are quite technical \cite{jslc}, so that we just review the key
results here.

The discussion leading to (\ref{laec}) already suggests that the critical
behavior of the driven system is distinct from its equilibrium counterpart:
the upper critical dimension is shifted to $d_{c}=5$, and parallel and
transverse wave vectors scale with different exponents. Anticipating
renormalization, we reformulate their scaling as $|{\bf k}_{\Vert }|\sim |%
{\bf k}_{\bot }|^{1+\Delta }$, introducing the {\em strong anisotropy }%
exponent $\Delta $. To illustrate its importance, let us consider the wave
vector scaling for the equilibrium Ising model. For isotropic exchange
interactions, coarse-graining results in the usual Landau-Ginzburg
Hamiltonian with gradient term $(\nabla \varphi )^{2}$. Clearly, this cannot
lead to anything but $|{\bf k}_{\Vert }|\sim |{\bf k}_{\bot }|$. The {\em %
only} effect of anisotropies in the microscopic couplings is to give rise to
different {\em amplitudes}, so that $\Delta $ remains zero. We refer to this
situation as {\em weak} anisotropy, in contrast to {\em strong }anisotropy
where $\Delta \neq 0$. Examples of the latter in equilibrium models include,
e.g., Lifshitz points or structural phase transitions \cite{lifshitz}.

For any system with strong anisotropy, irrespective of its universality
class, the renormalization group predicts the general scaling form of, e.g.,
the dynamic structure factor near criticality: 
\begin{equation}
S({\bf k},t;\tau _{\bot })=\mu ^{-2+\eta }S(k_{\Vert }/\mu ^{1+\Delta },{\bf %
k}_{\perp }/\mu ,t\mu ^{z};\tau _{\perp }/\mu ^{1/\nu })  \label{SF}
\end{equation}

Here, $\mu $ is just a scaling factor. Eqn. (\ref{SF}) can be viewed as a 
{\em definition} of the critical exponents $\nu $, $z$, $\eta $ and $\Delta $%
. The appearance of the latter is of course consistent with our earlier
discussion. Different universality classes are distinguished by the
characteristic values of these exponents, expressed, e.g., through their $%
\epsilon $-expansions. For our model, all exponents, {\em except} $\Delta $,
take their mean-field values: $\nu =\frac{1}{2}$, $z=4$, $\eta =0$ while $%
\Delta =1+\frac{\epsilon }{3}$. A separate analysis yields the order
parameter exponent $\beta =\frac{1}{2}$. Note that all of these equalities
are {\em exact}, i.e., all higher order terms in the $\epsilon $-expansion
vanish!

Care must be taken on two fronts, both associated with the presence of
strong anisotropy, when comparing these exponents to Monte Carlo data.
First, in the equilibrium Ising model, the {\em same} exponent $\eta $
characterizes the divergence of the critical structure factor near the
origin in ${\bf k}$-space, $S({\bf k},t=0;\tau _{\perp }=0)\sim |{\bf k}%
|^{-(2-\eta )}$ for ${\bf k}\rightarrow 0$, and the power law decay of its
Fourier transform, the two-point correlations, $G({\bf r},t=0;\tau _{\perp
}=0)\sim |{\bf r}|^{-(d-2+\eta )}$ as $|{\bf r}|\rightarrow \infty $. In
contrast, {\em four} $\eta $-like exponents are needed in the driven case!
Fortunately, scaling laws relate them to $\eta $ and $\Delta $. For example,
we can {\em define} $\eta _{\bot }$ via $S(k_{\Vert }=0,{\bf k}_{\perp
},t=0;\tau _{\perp }=0)\sim |{\bf k}_{\perp }|^{-(2-\eta _{\bot })}$, for $|%
{\bf k}_{\perp }|\rightarrow 0$. Using (\ref{SF}), we read off the simple
result $\eta _{\bot }=\eta $. Similarly, we introduce $\eta _{\Vert
}^{\prime }$ through the relation $G(r_{\Vert },{\bf r}_{\bot }={\bf 0}%
,t=0;\tau _{\perp }=0)\sim r_{\Vert }^{-(d-2+\eta _{\Vert }^{\prime })}$,
for $r_{\Vert }\rightarrow \infty $. Keeping track of $\Delta $ in the
Fourier transform, we obtain the less trivial scaling law $\eta _{\Vert
}^{\prime }=\frac{\eta -\Delta (d-3)}{1+\Delta }$. Since most simulations
have focused on two-dimensional systems, we set $\epsilon =3$, predicting $%
\Delta =2$ and $\eta _{\Vert }^{\prime }=2/3$. Monte Carlo data for the pair
correlation along the drive beautifully display the expected $r_{\Vert
}^{-2/3}$ decay \cite{LR}. Numerical evidence for the strong anisotropy
exponent is somewhat more indirect, extracted from an anisotropic finite
size scaling analysis \cite{KTLJSW}. Convincing data collapse is obtained,
using anisotropic systems of size $L_{\Vert }\times L_{\bot }$ with fixed
``aspect'' ratio $L_{\Vert }/L_{\bot }^{1+\Delta }$, for the theoretically
predicted values of the exponents.

It is intriguing that the signals of continuous phase transitions in or near
equilibrium, namely, a diverging length scale, scale invariance and
universal behavior, also mark such transitions in far-from-equilibrium
scenarios.

\section{Some Recent Developments}

Beyond the topics discussed in the previous section, we may arbitrarily name
three other ``levels'' of non-equilibrium steady-state systems. The first
are associated with the KLS model itself, including fascinating results on
higher correlations \cite{3pf}, failure of the Cahn-Hilliard approach to
coarsening dynamics \cite{CH}, models with {\em shifted periodic }and/or 
{\em open }boundary conditions \cite{SPBCOBC}, and systems with AC or random
drives \cite{RDDS,BS}. Topics in the next ``level'' involve various
generalizations of KLS, such as other interactions \cite{LSZ-Szabo,Aniso},
multi-layer systems, multi-species models, and systems with quenched
impurities \cite{qImp}. Further ``afield'' is a wide range of driven
systems, e.g., surface growth, electrophoresis and sedimentation, granular
and traffic flow, biological and geological systems, etc. In these brief
lecture notes, we present two topics at the ``intermediate level'': a driven
bi-layer system and a model with two species.

\subsection{Phase Transitions in a Bi-layer Lattice Gas}

The physical motivation for considering multi-layered systems may be traced
to intercalated compounds \cite{dressel}. The process of intercalation,
where foreign atoms diffuse into a layered host material, is well suited for
modeling by a driven lattice gas of several layers. In both physical systems
and Monte Carlo simulations of a model with realistic parameters \cite
{carlow}, finger formation has been observed. Both are transient phenomena,
so that we might ask: are there any novel phenomena in the driven {\em %
steady states}? On the theoretical front, there are two motivations. To
study critical behavior of the single layer KLS model, it is necessary to
set the overall density at 1/2, with the consequence that two interfaces
develop below criticality. These interfaces seriously complicate the
analysis, since their critical behavior is quite distinct from the bulk. One
way to avoid these difficulties is to consider a bi-layer system \cite{Mon},
with {\em no inter-layer interactions}. Through particle exchange between
the layers, the system may order into a dense and a hole-rich layer, neither
of which has interfaces. Certainly, this is expected (and observed) in the
equilibrium case, where the steady state is just two Ising models, decoupled
except for the overall conservation law. The other is using multilayer
models as ``interpolations'' from two- to three-dimensional systems \cite{am}%
. Motivations aside, what the simulations reveal is entirely unexpected.

In the simplest generalization, a bi-layer system is considered as a fully
periodic $L\times L\times 2$ model. The first simulations were carried out
on the case with zero inter-layer interactions \cite{am}. Therefore, within
each layer, all aspects are identical to the KLS model. Particle exchanges 
{\em across} the layers, unaffected by the drive, are updated according to
the local energetics alone, so that these jump rates are identical to a
system in equilibrium. The overall particle density is fixed at 1/2. In the
absence of $E$, there would be a single second order transition at the
Onsager temperature. Since the inter-layer coupling is zero, phase
segregation at low temperatures will occur across the layers, i.e., the
ordered phase is characterized by homogeneous layers of different densities
(opposite magnetizations, in spin language). Having no interfaces, the
free energy of the system is clearly lowest in such a state. At $T=0$, one
layer will be full and the other empty, so that this state will be labeled
by FE.

Intriguingly, when $E$ is turned on, {\em two} transitions appear \cite{am}%
! As $T$ is lowered, the homogeneous, disordered (D) phase first gives way
to a state with strips in both layers, reminiscent of two entirely
unrelated, yet aligned, single layer driven systems. We will refer to this
state as the strip phase (S). As $T$ is lowered further, a first order
transition takes the S phase into the FE state. Why the S state should
interpose between the D and FE states was not understood. This puzzle,
together with the presence of interlayer couplings in intercalated
compounds, motivated our study of the {\em interacting} bi-layer driven
system \cite{colin}. Within this wider context, the presence of two
transitions is no longer a total mystery. On the other hand, this study
reveals several unexpected features, leading to interesting new questions.

Our system consists of two $L^2$ Ising models with attractive interparticle
interactions of {\em strength unity}. Arranged in a bi-layer structure, the 
{\em inter-layer} interaction is specified by $J$. Thus, the ``internal''
Hamiltonian is just ${\cal H}\equiv -4\sum nn^{\prime }-4J\sum nn^{\prime
\prime },$ where the first sum is over nearest neighbors {\em within} a
given layer while $n$ and $n^{\prime \prime }$ differ only by the layer
index. The external drive, $E$, is imposed through the jump rates, as in
Section II. For simulations, we kept the overall density at 1/2 and chose $%
J\in $ $[-10,10]$. Note that negative $J$'s are especially appropriate for
intercalated materials \cite{dressel,carlow}. The eventual goal of such a
study is to map out the phase diagram in the $T$-$J$-$E$ space. So far, we
have data for only three (positive) values of $E$. Not expecting further
surprises, we believe that we have uncovered the main features of the phase
diagram.

To begin, let us point out various features in the equilibrium case. Here,
the $J>0$ and the $J<0$ systems can be mapped into each other by a gauge
transformation. Thus, in the thermodynamic limit, the phase diagram in the $T
$-$J$ plane is exactly symmetric. However, for the lattice gas constrained
by a conservation law, the low temperature states of these two systems are
not the same, being S or FE, respectively. While the D-S or D-FE transitions
are second order (shown as a line in Fig. 3), the S-FE line will be first
order in nature ($T$ axis from 0 to 1 in Fig. 3). Due to the presence of
interfaces, the FE domain includes the $J\equiv 0$ line. For the same
reason, it ``intrudes'' into the $J>0$ half plane by $O(1/L)$, for finite
systems. Of course, on the $T$ axis lies a pair of decoupled Ising systems,
so that $T_{c}(J$=$0)$ assumes the Onsager value, as $L\to \infty $. In
general, $T_{c}(J)$ is not known exactly, though $T_{c}(J$=$\pm \infty )$ is
precisely $2T_{c}(J$=$0)$, since every configuration of the two layers can
be mapped into one in the usual Ising model. The lesson from equilibrium is
now clear: the S state is as generic as the FE state in this wider context.
As we will see below, the presence of two transitions does not represent a
qualitative change from the equilibrium phase structure.

\begin{figure}[htbp]
\hspace*{1.5cm} 
\epsfxsize=4in \epsfbox{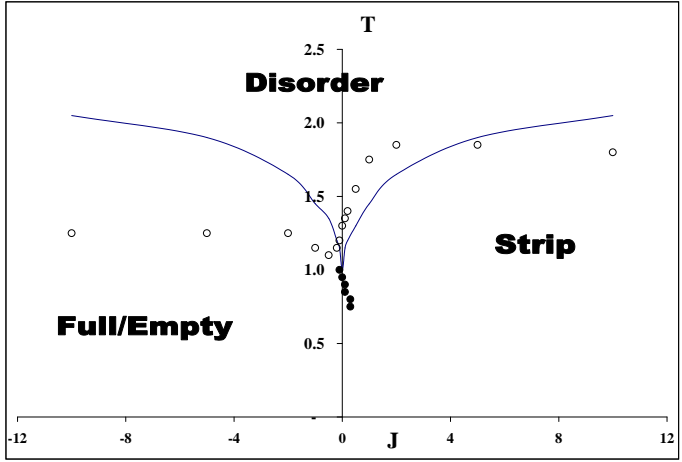} \vspace*{-1cm}
\caption{Phase diagrams for the bi-layer Ising lattice gas. Solid lines are
second order transitions in the equilibrium case. Open/solid circles are
continuous/discontinous transitions in the driven case.}
\end{figure}

Turning to driven systems, we can no longer expect a symmetric phase
diagram, since the drive breaks the Ising symmetry. The data for $E=25$ are
shown in Fig. 3, in which the first/second order transitions are marked as
solid/open circles. We see that the only effect of the drive is to shift the
phase boundaries. No new phases appear while the nature of the transitions
remains unchanged. However, there are remarkable features. One is the {\em %
lowering} of the critical temperature for large $\mid \!J\mid $. Given that $%
T_{c}(E)>T_{c}(0)$ in the KLS model, it is quite unexpected that $T_{c}
(|J| \gg 1,E\gg 1)$ is {\em less }%
than its equilibrium counterpart. Perhaps more notable is the presence of a
small region in the $J<0$ half plane, in which an S-phase is stable. Since
particle-rich strips lie on top of each other, such a phase could not exist
if either energy or entropy were to dominate the steady state. Concerned
that its presence might be a finite size effect, we performed simulations
using $L$'s up to $100$, with $J=-0.1$ and $T\in [1.00,1.20]$. In all cases,
the S-phase prevailed, leading us to conjecture that this region exists even
in the thermodynamic limit. On the other hand, for lower $T$, the FE-phase
penetrates into the $J>0$ half plane, as in equilibrium cases. Energetics
seem to take the upper hand here. Though we believe that, as $L\to \infty $,
this part of the phase boundary will collapse onto the $J=0$ line, we have
not checked the finite size effects explicitly. Of course, these
``intrusions'' into the positive/negative $J$ region by the FE/S-phase are
responsible for the appearance of {\em two }transitions in the studies of
the $J\equiv 0$ model \cite{am}.

It would clearly be useful to develop some intuitive arguments which can
``predict'' these qualitative modification of the equilibrium phase
structure. Since the usual energy-entropy considerations appear to fail, our
attempt \cite{colin} is based on the competition between suppression of
short-range correlations \cite{short range} and enhancement of long-range
ones \cite{LR}, as a result of driving. Let us focus on the two-point
correlations in the disordered phase with $T$ being lowered from $\infty $
and see how they are affected by the drive. On the one hand, the {\em %
nearest-neighbor} correlations are found to be suppressed by $E$, consistent
with the picture that the drive acts as an extra noise in breaking bonds.
Taken alone, this suppression would lead to the lowering of the critical
temperature, as pointed out in Section II. On the other hand, as we showed
above, the drive changes significantly the large distance behavior of $G$,
from an exponential to a power law. Further, the amplitude is positive
(negative) for correlations parallel (transverse) to $E$. Both the positive
longitudinal correlations and the negative transverse ones should help the
process of ordering into strips parallel to the field. In other words, the
enhanced long-range parts favor the S-phase. Taken alone, we expect this
effect to increase $T_{c}(E)$. Evidently, for the single layer case, the
latter effect ``wins''. For a bi-layer system, we need to take into account
cross-layer correlations, which are {\em necessarily} short-ranged.

Focusing first on the $J=0$ case, where short-range effects due to
cross-layer interactions are absent, we are led to $T_c(0,E)/T_c(0,0)>1$.
Indeed, this ratio is comparable to that in the single-layer case. Next, we
consider systems with positive $J$. Without the drive, $T_c(J,0)$ is of
course enhanced over $T_c(0,0)$. For non-vanishing drive, it is not possible
to track the competition between the short- and long-range properties of the
transverse correlations. Evidently, for small $J$, the long-range part still
dominates, so that $T_c(J,E)/T_c(J,0)>1.$ However, for $J\gg J_0$ , the
presence of $E$ effectively lowers $J$, since the latter is associated with
only short-range correlations. Since a lower effective $J$ naturally gives
rise to a lower $T_c$ , we would ``predict'' that $T_c(J,E)/T_c(J,0)$ could
decrease considerably as $J$ increases. From Fig.3, we see that $%
T_c(J,E)<T_c(J,0)$ for $J\ge 5!$ The interplay of the competing effects is
so subtle that either can dominate, in different regions of the phase
diagram. Finally, we turn to the $J<0$ case. Here, the two effects tend to
co-operate rather than compete, since the long-range parts favor the S-phase
over the FE-phase. As a result, the domain of FE is smaller everywhere. In
particular, note that the $J<0$ branch of $T_c(J,E)$ is significantly lower
than the $J>0$ branch. The small region of the S-phase can be similarly
understood, at least at the qualitative level. In this picture, we may argue
that the bicritical point and its trailing first order line should be
``driven'' to the $J<0$ half-plane.

To end this subsection, we point out a few other interesting features. The
order parameters of the S and FE phases are conserved and non-conserved,
respectively. Based on symmetry arguments, we may expect that the critical
properties of the two second order branches will fall into different
universality classes. In contrast, the gauge symmetry forces {\em static
equilibrium} properties to be identical along these two branches \cite{shz}.
Since the drive breaks this symmetry, we can only speculate that the D-S
transitions belong to the same class as the single layer, KLS model \cite
{jslc} while the D-FE ones should remain in the equilibrium Ising class \cite
{GYH}. Since these two classes are distinct, we can also expect a rich
crossover structure near the bi-critical point. Assuming these conjectures
are confirmed, we can truly marvel at the novelties a driving field can
bring.

\subsection{Biased Diffusion of Two Species}

If the particles in the KLS model are considered ``charged'', it is natural
to explore the effects of having both positive and negative charges. Defined
on a fully periodic $L_{x}\times L_{y}$ square lattice, a configuration of
such a ``two species'' model can be characterized by two local occupation
variables, $n_{{\bf r}}^{+}$ and $n_{{\bf r}}^{-}$, which equal $1$ ($0$),
if a positive or negative particle is present (absent) at site ${\bf r\equiv 
}(x,y)$. In spin language, this corresponds to a spin-$1$ model, and novel
behavior is to be expected, since we have altered the internal symmetries of
the local order parameter. For simplicity, we will restrict ourselves to
zero total charge, $Q\equiv \sum_{{\bf r}}(n_{{\bf r}}^{+}-n_{{\bf r}%
}^{-})=0 $. Pursuing the analogy to electric charges, positive (negative)
charges move preferentially along (against) the drive $E$ which is directed
along the $-y$ axis. As a first step, we assume that there are no
interparticle interactions apart from an excluded volume constraint. Thus,
our model can be viewed as the high-temperature, large $E$ limit of a more
complicated interacting system. To model biased diffusion, a particle with
charge $q=\pm 1$ jumps onto a nearest-neighbor empty site according to $\min
[1,e^{-qE\,\delta y}]$, where $\delta y=0,\pm 1$ is the change in the $y$%
-coordinate of the particle. To allow charge exchange between neighboring
particles, two nearest-neighbor sites carrying opposite charge may exchange
their content with probability $\gamma \min [1,e^{-qE\,\delta y}]$. Now, $%
\delta y$ is the change in the $y$-coordinate of the {\em positive}
particle. The parameter $\gamma $ sets the ratio of the characteristic time
scales controlling, respectively, charge exchange and diffusion.

Physical motivations for this model come from various directions: fast ionic
conductors with several species of mobile charges \cite{FIC}, electric
breakdown of water-in-oil microemulsions \cite{AN}, and gel electrophoresis
of charged polymers \cite{gel el}. It can also be interpreted as a simple
model for some traffic or granular flows \cite{traffic,gran}.

Summarizing our simulation data, we first map out the phase diagram of the
model, in the space spanned by $E$, the total mass density $\bar{m}\equiv
\sum_{{\bf r}}(n_{{\bf r}}^{+}+n_{{\bf r}}^{-})/(L_{x}L_{y})$, and $\gamma $ 
\cite{ksz1}. Focusing on small $\gamma $, the diffusive dynamics is limited
by the excluded volume constraint. For sufficiently small $E$ and $\bar{m}$,
typical configurations are disordered, characterized by homogeneous charge
and mass densities and fairly large currents. Nevertheless, this phase is
highly nontrivial, supporting anomalous two-point correlations reminiscent
of the KLS model: In generic directions, the familiar $r^{-d}$ decay
prevails, with the {\em remarkable} exception of the field direction, where
a {\em novel} $r_{\Vert }^{-(d+1)/2}$ dominates \cite{ksz2}. With increasing 
$E$ or $\bar{m}$, the tendency of the particles to impede one another
becomes more pronounced, until a phase transition into a spatially
inhomogeneous ``ordered'' phase occurs. Beyond the transition line $E_{c}($ $%
\bar{m},\gamma )$, typical configurations are charge-segregated, consisting
of a strip of mostly positive charges ``floating'' on a similar strip of
negative charges, surrounded by a background of holes.

To explore these transitions more quantitatively, we need to define a
suitable order parameter. Taking the Fourier transforms of the local {\em %
hole} and {\em charge} densities, $\tilde{\phi}(k_{\bot },k_{\Vert })\equiv 
\frac{1}{L_{x}L_{y}}\sum_{{\bf r}}[1-(n_{{\bf r}}^{+}+n_{{\bf r}}^{-})]\exp
\{ik_{\bot }x+ik_{\Vert }y\}$ and $\tilde{\psi}(k_{\bot },k_{\Vert })\equiv 
\frac{1}{L_{x}L_{y}}\sum_{{\bf r}}[n_{{\bf r}}^{+}-n_{{\bf r}}^{-}]\exp
\{ik_{\bot }x+ik_{\Vert }y\}$, we select the components $\Phi \equiv |\tilde{%
\phi}(0,\frac{2\pi }{L_{y }})|$ and $\Psi \equiv |\tilde{\psi}(0,\frac{2\pi 
}{L_{y }})|$ since these are most sensitive to ordering into a single
transverse strip. Naturally, we choose their averages $<\Phi> $
and $<\Psi>$ as our order
parameters. To distinguish first from second order transitions, we also
measure their fluctuations and histograms. The resulting phase diagram is
shown in Fig. 4. Typically, hysteresis loops in $<\Phi>$ ,
$<\Psi> $ and the average current
are observed only for $\bar{m}<\bar{m}_{o}(\gamma )$, indicating that the
transitions are first order in this regime. On the other hand, fluctuations
develop sharp peaks for larger mass densities, signalling continuous
transitions. Larger values of $\gamma $ favor the disordered phase, so that
the whole transition line shifts to higher $E$, the region between the
spinodals narrows, and $\bar{m}_{o}(\gamma )$ moves to higher mass
densities. Thus, we observe a surface of first order transitions, separated
from a surface of continuous ones by a line of multicritical points: $E_{c}(%
\bar{m}_{o}(\gamma ),\gamma )$.
\begin{figure}[htbp]
\hspace*{1.5cm} 
\epsfxsize=4in \epsfbox{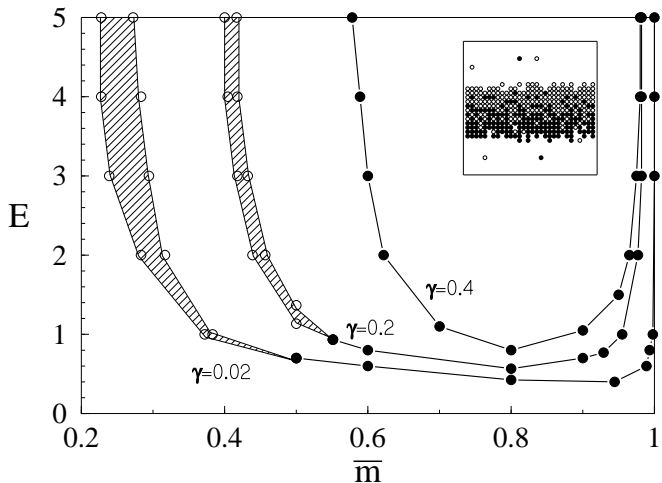} \vspace*{-1cm}
\caption{Phase diagram for the driven two-species model. The system size is $%
30 \times 30$. The filled circles mark the lines of continuous transitions,
while the open circles denote the spinodal lines associated with first
order transitions. The inset shows a typical ordered configuration at $\bar{m}%
=0.40$, $E=3.00$ and $\gamma=0.02$ with the open (filled) circles
representing positive (negative) charges.}
\end{figure}

While it is not surprising that the disordered phase is stable for large $E$
provided the mass density is sufficiently small, it may appear rather
counterintuitive that it should also dominate the $\bar{m}\lesssim 1$
region. However, setting $\bar{m}\equiv 1$ eliminates every hole in the
system so that the dynamics is carried entirely by the charge exchange
mechanism. Relabelling positive charges as ``particles'' and negative ones
as ``holes'', the model becomes equivalent to a noninteracting (i.e., $J=0$)
KLS model. Also termed the asymmetric simple exclusion process (ASEP) in the
literature, its steady-state solution is exactly known \cite{Spitzer} to be
homogeneous. Thus, for all $\gamma >0$, our two-species model remains
disordered along the entire $\bar{m}\equiv 1$ line. For sufficiently large $%
\gamma $, the data indicate quite clearly that a {\em finite} region of
disorder persists, for any $E$, just {\em below} complete filling. For
smaller $\gamma $, however, it is less obvious whether such a region exists
in the thermodynamic limit, since even a single hole can suffice to induce
spatial inhomogeneities in a {\em finite} system: Acting as a catalyst for
the charge segregation process, the hole creates a strip of predominantly
positive charge, separated by a rather sharp interface from a similar,
negatively charged domain, located ``downstream''. The charge exchange
mechanism tends to remix the charges but has little effect for small $\gamma 
$. Eventually, the hole ends up trapped in the interfacial region! Finally,
we turn to larger values of $\gamma $, specifically, $\gamma \gtrsim 0.62$:
here, the charge exchange mechanism dominates over the excluded volume
constraint and suppresses the ordered phase completely.

So far, our discussion has mostly drawn upon Monte Carlo results. It is
gratifying, however, that the same qualitative picture emerges from a
mean-field theory, based upon a set of equations of motion for the local
densities. We briefly summarize the analytic route. Since the numbers of
both positive and negative charges are conserved, we begin with a continuity
equation for the coarse-grained densities $\rho ^{\pm }({\bf r},t)$: $%
\partial _{t}\rho ^{\pm }+\nabla {\bf j}^{\pm }=0$. For simplicity, we focus
on the case $\gamma =0$, i.e., particle-hole exchanges, first. As for the
KLS model, ${\bf j}^{\pm }$ can be written as the sum of a diffusive piece
and an Ohmic term, ${\bf j}^{\pm }=\lambda ^{\pm } \{-\nabla \frac{\delta {\cal H}}{%
\delta \rho ^{\pm }}_{|\rho ^{\mp }}\pm {\cal E}{\bf \hat{y}}\}$. The
density-dependent mobility $\lambda ^{\pm } $ must vanish with both $\rho ^{\pm }$
and the local hole density, $\varphi \equiv 1-(\rho ^{+}+\rho ^{-})$, i.e.,
$\lambda ^{\pm } = \rho ^{\pm }\varphi $. The
``Hamiltonian'' ${\cal H}=\{\rho ^{+}\ln \rho ^{+}+\rho ^{-}\ln \rho
^{-}+\varphi \ln \varphi \}$ is just the mixing entropy associated with
distributing $L_{x}L_{y}\rho ^{+}$ positive and $L_{x}L_{y}\rho ^{-}$
negative charges over a lattice of $L_{x}L_{y}$ sites. Note that the
functional derivative $\frac{\delta {\cal H}}{\delta \rho ^{\pm }}$ is taken
at fixed $\rho ^{\mp }$ since we are focusing on particle-hole exchanges
here. To model the charge exchange mechanism, we simply add a similar term
to ${\bf j}^{\pm }$, namely, ${\bf j}^{\pm }=\lambda \{-\nabla \frac{\delta 
{\cal H}}{\delta \rho ^{\pm }}\pm {\cal E}{\bf \hat{y}}\}+\gamma \lambda
^{\prime }\{-   \nabla \frac{\delta {\cal H}}{\delta \rho ^{\pm }}_{|\varphi
}\pm {\cal E}{\bf \hat{y}}\}$ where $\lambda ^{\prime } = \rho ^{+} \rho ^{-} $ 
vanishes with both $%
\rho ^{+}$ and $\rho ^{-}$, and the functional derivative is taken at fixed 
{\em hole} density. Expressing the resulting equations in the more
convenient variables $\varphi $ and charge density $\psi \equiv \rho
^{+}-\rho ^{-}$, we obtain 
\begin{eqnarray}
\partial _{t}\varphi  &=&\nabla \{\nabla \varphi +{\cal E}\varphi \psi {\bf 
\hat{y}}\}  \nonumber \\
\partial _{t}\psi  &=&\nabla \{\gamma \nabla \psi +(1-\gamma )[\varphi
\nabla \psi -\psi \nabla \varphi ]
-{\cal E}\varphi (1-\varphi ){\bf \hat{y}}  \label{2sp}
\\
&&-\frac{\gamma }{2}{\cal E}[(1-\varphi )^{2}-\psi ^{2}]{\bf \hat{y}}\} 
\nonumber
\end{eqnarray}
These equations have to be supplemented by periodic boundary conditions and
constraints on total mass and charge, $(1-\bar{m})L_{x}L_{y}=\int
d^{d}r\varphi ({\bf r},t)$ and $0=\int d^{d}r\psi ({\bf r},t)$. Also, to be
precise, all Laplacians $\nabla ^{2}$ should be given the appropriate
anisotropic interpretation.

Using (\ref{2sp}) as our starting point, we seek to recapture the major
features of the phase diagram. First, the presence of two phases, uniform
vs. spatially structured, is reflected in the existence of both homogeneous
and inhomogeneous steady-state solutions of (\ref{2sp}). The existence of
the former is evident, due to the conservation law. Anticipating homogeneity
in the transverse directions, we seek the latter in the form $\varphi (y)$, $%
\psi (y)$. Integrating (\ref{2sp}) once, the stationary mass and charge
currents appear as natural integration constants. Since the numbers of
positive and negative charges are equal, the former must be zero, leaving us
with the latter: $J$. The first equation allows us to eliminate $\psi $ in
favor of $\varphi $, and, with $u\equiv \sqrt{1+\frac{\gamma }{(1-\gamma
)\varphi }}$ as the new variable, we can reduce the second equation to
potential form, $u^{\prime \prime }=-\frac{dV(u)}{du}$. The potential $V(u)$,
given explicitly in \cite{ksz1},
exhibits a minimum for a range of $J$'s, so that inhomogeneous, periodic
solutions of (\ref{2sp}) exist. We note in passing that these may even be
found explicitly, provided $\gamma =0$ \cite{vzs}. Otherwise, numerical
integration is of course always possible, yielding rather impressive
agreement with simulation profiles. Interestingly, our equations of motion
predict that ${\cal E}$ enters only through the {\em scaling variable }$%
{\cal E}L_{y}$. If plotted accordingly, our simulation data confirm
this scaling by collapsing rather convincingly, at least within their error
bars.

However, the mere {\em existence} of two types of steady-state solutions is
not sufficient to provide evidence for a phase transition. Therefore, we
seek instabilities of these solutions, as the control parameters ${\cal E}$, 
$\bar{m}$ and $\gamma $ are varied. Performing a linear stability analysis,
we find that a {\em homogeneous} solution with mass $\bar{m}$ becomes
unstable if the drive ${\cal E}$ exceeds a threshold value ${\cal E}_{H}(%
\bar{m},\gamma )=\frac{2\pi }{L_{y}}\sqrt{\frac{1-\bar{m}+\gamma \bar{m}%
}{(1-\bar{m})[(2-\gamma )\bar{m}-1]}}$. The most relevant perturbation is
associated with the smallest wave vector in the parallel direction, $({\bf 0,%
}2\pi /L_{y })$. A simple analysis of ${\cal E}_{H}(\bar{m},\gamma )$
shows that this instability can only occur within the interval $\frac{1}{%
2-\gamma }<\bar{m}<1$, so that, in particular, the homogeneous phase is
always stable for $\gamma \geqslant 1$. Clearly, we cannot identify this
mean-field stability boundary with the true transition line: we have not
considered the locus of instabilities associated with {\em inhomogeneous}
solutions, and fluctuations have been neglected throughout. However, it is
quite remarkable how well it mirrors the qualitative shape of the phase
diagram.

Finally, let us consider the {\em order} of the transitions. In principle,
two routes can be pursued here. One of these, namely, the computation of the
stability boundary of the {\em inhomogeneous} phase, ${\cal E}_{I}(\bar{m}%
,\gamma )$, is rather subtle, involving three Goldstone modes \cite{inh stab}%
. Once ${\cal E}_{I}(\bar{m},\gamma )$ is known, values $(\bar{m},\gamma )$
for which ${\cal E}_{I}$ and ${\cal E}_{H}$ coincide can be identified as
loci of continuous transitions; otherwise, ${\cal E}_{H}$ and ${\cal E}_{I}$
mark the ``spinodals'', near a first order transition, where the
homogeneous/inhomogeneous phases become linearly unstable. Alternatively,
the adiabatic elimination of the fast modes results in an effective equation
of motion for the slow ones which can then be analyzed. Since the technical
details of this approach can be found in \cite{ksz1}, we need only focus on
the result which combines the expected with the surprising. Letting $%
M(q_{\bot },t)$ denote the {\em complex} amplitude of the unique slow mode,
associated with the band of wave vectors $(q_{\bot }{\bf ,}2\pi /L_{y})$%
, its equation of motion takes the Ginzburg-Landau form: $\partial
_{t}M=-\{(\tau +q_{\bot }^{2})M+gM|M|^{2}+O(M^{5})\}$. Here, $\tau $ is the
soft eigenvalue, which vanishes on the stability boundary and $g=g({\cal E},%
\bar{m},\gamma ,L_{y})$ is a rather complicated function. As in
standard Landau theory, the sign of $g$ determines the order of the
transition: if positive, the transition is continuous while negative $g$
signals a first order one. Since $\tau \simeq 0$, we can set ${\cal E}={\cal %
E}_{H}$ and discuss $g$ on the stability boundary itself: being positive for 
$\bar{m}\lesssim 1$, it has a unique zero at a critical $\bar{m}_{o}(\gamma )
$, below which it becomes negative. Since $\bar{m}_{o}(\gamma )$ increases
with $\gamma $, we recover the qualitative behavior of the multicritical
line observed in the simulations. The {\em surprising} aspect of this
equation of motion, however, resides in the fact that $M$ is O(2)-symmetric
and $q_{\bot }$ spans just a single spatial dimension. Given these
symmetries, the Mermin-Wagner theorem \cite{mw} should forbid the existence
of long-range order! One might hope that a careful analysis of finite-size
effects in this system would contribute to the disentanglement of this
puzzle.

We conclude with two comments. First, we noted earlier that the case $\bar{m}%
\equiv 1$ is exactly soluble. There are, in fact, two other surfaces in the
phase diagram, corresponding to $\gamma =1$ and $\gamma =2$, for which
certain distributions are exactly known. Setting $\gamma =1$ implies that
the rates for particle-hole and particle-particle exchanges become equal, so
that any given particle, e.g., a positive charge, cannot distinguish between
negative charges and holes. Thus, it experiences biased diffusion,
equivalent to the non-interacting KLS model. Accordingly, the {\em marginal }%
steady-state distribution of the occupation numbers of {\em one} species is
uniform, i.e., $P\left[ \{n_{{\bf r}}^{\pm }\}\right] =\sum_{\{n_{{\bf r}%
}^{\mp }\}}P\left[ \{n_{{\bf r}}^{+},n_{{\bf r}}^{-}\}\right] \propto 1$,
and observables pertaining to a single species are trivial. For example, the
two-point correlation functions for equal charges, $<n_{{\bf r}%
}^{+}n_{{\bf 0}}^{+}>$ and $<n_{{\bf r}}^{-}n_{%
{\bf 0}}^{-}>$, vanish for ${\bf r\neq 0}$. The {\em full}{\bf %
\ }distribution, $P\left[ \{n_{{\bf r}}^{+},n_{{\bf r}}^{-}\}\right] $,
however, is nontrivial, so that, e.g., the two-point function for {\em %
opposite} charges, $<n_{{\bf r}}^{+}n_{{\bf 0}}^{-}>$, 
remains long-ranged \cite{ksz2}. A completely random system, marked by a
uniform $P\left[ \{n_{{\bf r}}^{+},n_{{\bf r}}^{-}\}\right] $ results only
if $\gamma =2$. Finally, we note that the one-dimensional version of our
model is exactly soluble by matrix methods \cite{Sandow}.

Second, it is natural to wonder about the consequences of having non-zero
charge. This problem has only been investigated for $\gamma =0$ \cite{Q},
but the findings are quite remarkable. The system still orders into a
charge-segregated strip, while supporting a {\em nonvanishing} mass current,
reflected in an overall drift. If, e.g., positive charges outnumber negative
ones, one might expect that the whole strip would drift in the field
direction - in analogy with American football, where the team with fewer
players tends to lose ground. In contrast, the strip wanders {\em against}
the field, following the preferred direction of the minority charge!

We should add that this model possesses several other intriguing properties,
e.g., stable configurations with nontrivial winding number 
(``barber poles'') \cite{barber} or
multiple-valued currents in the mean-field description \cite{vzs}. To
summarize, the remarkable richness of this
deceptively simple system is clearly amazing.

\section{Summary and Outlook}

We have presented, within the limited scope of these lecture notes, a brief
introduction to the statistical mechanics of driven diffusive systems and
some recent developments in this ever expanding field. Focusing only on the
prototype model and two of the simplest generalizations, we pointed out a
multitude of surprises, when we base our expectations on the experience with
equilibrium systems. While some of these, e.g., the generic presence of
singular correlations, are rather well understood, others, such as the
nature of ordering in two-species models, remain unresolved. Of course, the
holy grail of this whole field, namely, the fundamental understanding and
theoretical classification of non-equilibrium steady states, still beckons
at the distant horizon.

\section{Acknowledgments}

It is a pleasure to especially acknowledge our collaborators on the recent
work reported here: C.C. Hill, G. Korniss and K.-t Leung. Others are too
numerous to mention but no less deserving. We thank the IXth International
Summer School on Fundamental Problems in Statistical Mechanics, and
particularly H.K. Janssen and L. Sch{\"a}fer, for their hospitality. This
research was supported in part by grants from NATO, the SFB 237 of the
Deutsche Forschungsgemeinschaft and the US\ National Science Foundation
through the Division of Materials Research.

\end{document}